\documentclass[final,3p,number ]{elsarticle}
\usepackage{graphicx}
\usepackage{amsthm,amssymb,amsmath}
\usepackage[utf8]{inputenc}
\usepackage[T1]{fontenc}
\usepackage[update,prepend]{epstopdf}
\usepackage{siunitx} 
 
\journal{Wave Motion}
\begin{document}
\newcommand{\bra}[1]
{ \langle #1 |}
\newcommand{\ket}[1]
{| #1 \rangle}
 \newcommand{\bk}[2]
{\langle #1 | #2 \rangle}
  \newcommand{\modulc}[2]
{\left\lvert #1 \right\lvert ^{2}}
\newcommand{\module}[2]
{\left\lvert #1 \right\lvert }
\newcommand{\Imm}[1]
{\textrm{Im}\left [ #1 \right ]}
\newcommand{\Ree}[1]
{\textrm{Re} \left [ #1 \right ]}
\newcommand{\Dt}
 {\Delta_{\bot}}
\newcommand{\rp}{r^{\prime}}
\newcommand{\rc }{\textrm{reverberation chamber}}
\newcommand{\EM}{\textrm{electromagnetic}}
\newcommand{\modul}[2]{\left\lvert #1 \right\lvert }
\newcommand{\Nw}{ N_\textrm{Weyl}}
\begin{frontmatter}
\title{Universal behaviour of a wave chaos based electromagnetic reverberation chamber}
\begin{abstract} 
In this article, we present a numerical investigation of three-dimensional \EM \,  Sinai-like cavities. We computed around 600 eigenmodes for two different geometries: a parallelepipedic cavity with one  half-sphere on one wall and  a parallelepipedic cavity with one half-sphere  and two spherical caps on three adjacent walls. We show that the statistical requirements  of a well operating \rc \, are better satisfied in the more complex geometry without a mechanical mode-stirrer/tuner. This is to the fact that our proposed cavities exhibit spatial and spectral statistical behaviours very close to those predicted by random matrix theory. More specifically, we show that in the range of frequency corresponding to the first few hundred modes, the suppression of non-generic modes (regarding their spatial statistics) can be achieved by   reducing  drastically the amount of parallel walls.
 Finally, we compare the influence of  losses on the statistical complex response of the field inside a parallelepipedic and a chaotic cavity. We demonstrate that, in a chaotic cavity without any stirring process, the low frequency limit of a well operating \rc \, can be significantly reduced under the  usual values obtained in mode-stirred \rc s.
 \end{abstract}
 %\maketitle

\author[lpmc]{Jean-Baptiste Gros }
 \author[lpmc]{Olivier Legrand}
\ead{olivier.legrand@unice.fr}
\author[lpmc]{Fabrice Mortessagne}

\author[esy]{Elodie Richalot}
\author[esy]{Kamardine  Selemani}

\address[lpmc]{Universit{\'e} Nice Sophia Antipolis, CNRS, Laboratoire de Physique de la Mati{\`e}re Condens{\'e}e, UMR 7336, 06100 Nice, France}
\address[esy]{Universit{\'e} Paris-Est, ESYCOM, Marne-la-Vall{\'e}e, Cit{\'e} Descartes, 77454 Marne-la-Vall{\'e}e, France} 
 
 
\end{frontmatter}

\section{Introduction}
Electromagnetic reverberation chambers (RC) are nowadays commonly used for electromagnetic compatibility (EMC) applications \cite{Norme2003}. Thanks to the presence of a mechanical stirrer of irregular geometry (generally a rotating metallic object), electronic devices under test are submitted to an isotropic, statistically uniform and depolarised electromagnetic field. Those properties are satisfied  as long as the frequency is above the so-called \emph{lowest useable frequency} (LUF). 
Very often, this LUF is referred as the \emph{lowest overmoded frequency} \cite{Bruns2005,Hill1998a}. The overmoded condition suffers from the lack of a proper definition and is often associated to the concept a modal density threshold irrespective of the amount of losses \cite{hill2009electromagnetic}. Consequently, several definitions of the LUF can be found in the literature, that are not necessarily equivalent: the LUF can either be between three to six times the cutoff frequency $f_c$ of the fundamental mode, or defined as the frequency around which a hundred modes are counted above $f_c$ \emph{and} at which the modal density is greater than $1.5$ modes/MHz \cite{Bruns2005,Norme2003}. It is quite obvious that the latter definition strongly depends  on both the size of the chamber and on the importance of modal overlap related to losses \cite{Cozza2011} (either at walls or due to antennas). Moreover, as the LUF is understood as being the frequency above which the condition of a statistical uniformity of the field is achieved, the use of a stirrer is supposed to ensure the validity of the description of the electromagnetic field as  a random superposition of plane waves, which is coined the \emph{continuous plane wave-spectrum} hypothesis \cite{Hill1998,hill2009electromagnetic}. However, it has been acknowledged that even for $f>$LUF, the field may not be statistically uniformly distributed in the RC as a consequence of a bad stirring \cite{Arnaut2001,Lunden2007}. Thus, the behaviour of mode-stirred RCs proves to be highly non-universal, depending on the position and design of the stirrer \cite{Bruns2005,Hong2010}. To improve the properties of  an EM RC, the question of lowering the frequency range of well-operating RCs is primordial. However, Bruns et al. \cite{Bruns2005} have numerically established that all stirrers are equally ineffective below the conventional LUF whatever their shape, orientation, or electrical size.

The above mentioned statistical requirements of a well-stirred RC above the LUF closely correspond to the natural behaviour of a chaotic cavity \cite{Berry1977,wright2010_2,wright2010_4}. In this paper, we propose to investigate spectral and spatial statistical properties of three-dimensional (3D) chaotic cavities as a new paradigm for a reverberation chamber.
Indeed, in a chaotic cavity, generic modes (also called \emph{ergodic} modes \cite{Berry1977,wright2010_2,wright2010_4}) display Gaussian statistics of the fields. 
This statistical behaviour can be met at relatively low frequency, thereby suggesting that statistically isotropic fields and random polarisations can be obtained even in an unstirred cavity.
Quantum or Wave chaos has been a very active field of research for decades, concerned with the spectral and spatial asymptotic properties of wave systems whose ray counterpart is chaotic \cite{BGS,Bohigas1984,stockmann1999quantum}. Predictions of Random Matrix Theory (RMT) have been extensively verified, both numerically and experimentally in two-dimensional (2D) or pseudo-2D electromagnetic cavities \cite{McDonald1988,Stein1992,stockmann1999quantum,Alt1994,Pre_Doya} as well as in the 3D case \cite{Deus1995,Alt1997,Dembowski2002,Dietz2008,Dorr1998}.  
Previous works have already called for the similitude between the expected statistical properties a well-stirred RC and the intrinsic behaviour of individual \emph{ergodic} modes of a chaotic cavity, to propose strategies for improved operation of an EM RC \cite{Arnaut2001,Orjubin2009,Gradoni2011}.
We believe that, along with these studies, the results presented here have practical implications for an effective reduction of the lowest usable frequency (LUF) in chaotic RCs. 

In the present paper, we demonstrate that the description of the electromagnetic field as a continuous plane wave-spectrum \cite{Hill1998,hill2009electromagnetic} can only be justified at high frequencies where modal overlap is large, which is not necessarily the case near the LUF according to the commonly accepted definitions of the latter recalled above. 
Thus, the main reason why a conventional mode-stirred  RC can operate satisfactorily near the LUF  is that the presence of the stirrer makes it more like a  pseudo-integrable system, equivalent to a barrier 2D-billiard \cite{Hannay1999}, where many eigenmodes are similar to \emph{ergodic} modes of a chaotic cavity \cite{Bogomolny2004,Bogomolny2006,Dietz2008}. Nevertheless, even at high frequencies, pseudo-integrable cavities have a non negligible number of superscarmodes \cite{Bogomolny2006}, with non Gaussian field distributions. These unwanted features are clearly due to the non-universal spatial or spectral behaviour of pseudo-integrable cavities.
It is therefore the universality of chaotic cavities which will serve as the unifying thread of our investigations, keeping in mind that the operation of actual RCs is not restricted to high frequencies and that the modifications of geometry we propose should remain effective at a reduced cost.

In the following section, we introduce key concepts concerning spectral properties of chaotic wave systems, to provide diagnosis tools for RCs. More specifically, universal properties of spectral fluctuations are analysed through the nearest neighbour spacing distribution (NNSD) and the so-called Number Variance derived from the 2-level correlation function. We then investigate these quantities in 3D Sinai-like cavities to exemplify the influence of non-generic modes in departures from non-universal spectral behaviours in 3D chaotic cavities. We also illustrate how universal spectral properties are intimately connected to spatial properties of fields, namely the Gaussian distribution of amplitudes of field components for generic \emph{ergodic} modes.

In the final section, we study the effect of losses on the complex response of 3D RCs. Using the dyadic Green's function (DGF), in a perturbative approach, we study the spatial statistical distribution of fields near the LUF. By accounting for a small or moderate modal overlap we compare a classical rectangular RC without stirrer to a chaotic RC, demonstrating how the latter can fulfil the required statistical features of a well-stirred RC, concerning the spatial distribution of the field, without the use of a stirring process.

\section{Spectral and spatial statistics}
\subsection{Universal spectral statistics of chaotic cavities}
Since the Bohigas-Giannoni-Schmit conjecture \cite{BGS} concerning the universality of level fluctuations in chaotic quantum spectra, it is customary to analyse spectral fluctuations of chaotic cavities with the help of statistical tools introduced by RMT. These tools will be used to check the \emph{chaoticity} of the RCs we study. 

Starting from the modal density :
\begin{equation}
 \rho(f)=\sum_{i} \delta(f-f_i)
\end{equation}
$f_i$ denoting the eigenfrequencies of the cavity, one defines the counting function:
\begin{equation}
 N(f)=\int_0^f \rho(\nu) d\nu
\end{equation}
 whose smoothly varying part, denoted $\Nw$, is related to the geometry of the cavity through the Weyl's law \cite{Luk,Balian1977} which reads:
%\begin{equation}
%\Nw^{2D}(f)=\frac{A \pi }{c^2} f^2-\frac{P}{2c} f+const
%\end{equation} 
%in the case of a 2D scalar wave problem with Dirichlet boundary conditions, where $A$ is the area of the domain and $P$ its perimeter;
 \begin{equation}
 \Nw^{3D}(f)=a f^3+b f+const
 \end{equation}
 in the case of a 3D (vectorial) EM case and for perfectly conducting boundary conditions, where 
 \begin{align}
 a&= \frac {8\pi}{3c_0^3} V\\
b&=b_{curv}+b_{edges}\\
\textrm{with }  b_{curv}&= -\frac{4}{3\pi c_0}\int \frac{d\sigma}{R_m(\sigma)}\\
b_{edges}&=\sum_{e}\frac{1}{6\pi c_0}\int dl_e \frac{\left(\pi - \Omega(l_e) \right)\left(\pi - 5\Omega(l_e)\right )}{\Omega(l_e)} .
\end{align}
In the above formulae, $V$ is the volume of the cavity, $c_0$ is the speed of light, and  $R_m$ labels the mean radius of curvature over the surface $\sigma$  and $\Omega$ is the dihedral angle along the edges $e$. 
 Note that in the 3D case, there is no term proportional to  $f^2$ because the different  contributions of both  polarisations TE and TM  mutually cancel out.
 
 In order to obtain a statistical analysis of an actual sequence of eigenfrequencies \{$f_i$\},  independently of  the specific size of the resonator, the spectrum should be unfolded as follows.
 One defines the unfolded sequence $x_i=\Nw(f_i)$ and the sequence of spacings $s_i=x_{i+1}-x_i$, whose average is $\left\langle s_i\right\rangle=1$. Then, the fluctuations of the counting function $N$ around its smooth counterpart $\Nw$ can be characterised by the probability distribution $P(s)$  of normalised spacings  (the  NNSD) or, alternatively, by the Number Variance defined by \cite{Bohigas1984}:
 \begin{equation}
\label{sigma2}
\Sigma^2(L)=\left\langle (n_\gamma(L)-L)^2  \right\rangle_\gamma
\end{equation}
where $n_\gamma(L) $ is the number of modes $x_i \in \left[\gamma,\gamma+L\right]$ and  $\left\langle ... \right\rangle_\gamma$ is an average on $\gamma $.
If the resonant cavity is integrable, as in the case of a rectangular one, the expected NNSD follows the Poisson distribution \cite{Bohigas1984,stockmann1999quantum} : 
   \begin{equation}
    P_\textrm{Poisson}(s)= \exp(-s)
  \end{equation}
whereas in a fully chaotic cavity with  time reversal  symmetry, the Gaussian Orthogonal Ensemble (GOE) of RMT predicts a NNSD which is well approximated by the so-called Wigner distribution \cite{Bohigas1984,stockmann1999quantum} :
 \begin{equation}
    P_\textrm{Wigner}(s)=\frac{\pi}{2} s \exp(-\frac{\pi}{4}s^2)
\end{equation}
Both distributions correspond to special cases of the heuristic Weibull distribution \cite{Orjubin2007}
\begin{equation}
 P_\textrm{Weibull}(s)=\frac{\beta}{\alpha^\beta}s^{\beta-1}\exp(-\frac{s^\beta}{\alpha^\beta}) 
\end{equation}
where $\alpha=\left\langle s_i\right\rangle\Gamma^{-1}\left( 1+1/\beta\right)$. This distribution is also referred  to  the Brody distribution when $\left\langle s_i\right\rangle=1$ \citep{Orjubin2007,stockmann1999quantum}.
 Note that  $P_\textrm{Weibull}=P_\textrm{Wigner}$ if $\beta=2$ with $\left\langle s_i\right\rangle=1$ and $P_\textrm{Weibull}=P_\textrm{Poisson}$ if $\beta=1$ with $\left\langle s_i\right\rangle=1$. This distribution is generally used to fit numerical distributions since it continuously interpolates between Poisson and Wigner, thus enabling to quantify deviations  from  the universal behaviour  verified in  chaotic systems with time reversal symmetry. From the NNSD, one can derive a quantity more appropriate for comparison with experimental results: the cumulative distribution function $F(s)$, whic reads
\begin{equation}
\label{cdf}
F(s)=\int_0^s P(x) dx
\end{equation} 

In the following, all spectra we deal with will be restricted to sequences of about 500 frequencies. We should therefore consider the possibility that fitting our data with $P_\textrm{Weibull}$ will yield $\beta$ values which fluctuate.
For a large number  of  randomly generated samples, each composed of 500 spacings distributed according to $P_\textrm{Wigner}$, the  expected average value of parameter $\beta$ is equal to $\overline{\beta}=2.00$  with a standard deviation $\sigma_{\beta} = 0.07$. Thus typically, up to $2$ to $3$ standard deviations around $\overline{\beta}$, the $\beta$-values obtained from fitting our numerical data with $P_\textrm{Weibull}$ will be considered as compatible with the GOE predictions.
In a chaotic system, deviations from GOE can essentially be attributed either to the fact that the sequence of modes is not truly asymptotic and/or to the presence of non-generic modes (which become more and more exceptional at higher frequencies). These non-generic modes give rise to large scale fluctuations (generally with larger amplitudes than for a pure GOE spectrum) of the counting function $N$ with respect to its average $N_{av}$. In the following, such non-generic modes will be shown to be associated to continuous families of marginally stable periodic orbits of ray trajectories. These non-generic modes constitute the extension of bouncing-ball modes in 2D quantum billiards to the case of 3D EM cavities. In 2D quantum chaotic billiards \cite{Tanner1997,Backer1997}, bouncing ball modes concentrate on a phase-space region surrounding the continuous families of marginally stable periodic orbits, whereas ergodic modes uniformly spread out over the whole phase space.\\
 It should be noted here that $\Sigma^2$ is generally a quantity which proves to be more sensitive to the existence of non-generic modes than $P(s)$. So if $\Sigma^2$  is close to 
 $\Sigma^2_{GOE}$, one can expect that the chaotic system under study is in the asymptotic regime \cite{Alt1997}. 
 
\subsection{From regular to fully chaotic cavity}

\begin{figure}[h]
\center
\resizebox{10cm}{!}{\includegraphics{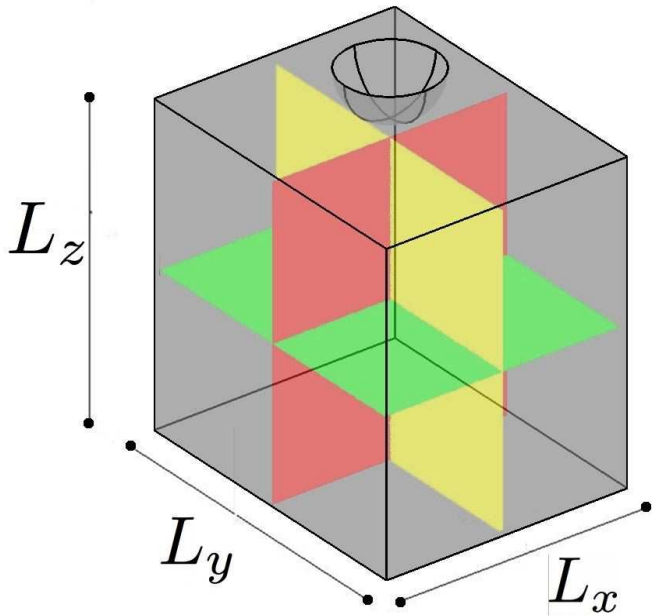}}
\caption{A simple modification of a regular RC with a single half-sphere on top. Planes associated  with tangential modes are shown in color.}
\label{schema}
\end{figure}

The ideal 3D chaotic cavity is given by a fully asymmetric room without any parallel walls and with defocusing parts (focusing parts can also be used, with restrictions regarding the centers of curvature \cite{Bunimovich1998,Berry1981}).  In this sense, using a parallelepipedic room to ensure a homogeneous repartition of energy is not optimal as recognized for decades by the community of room acoustics \citep{kuttruff2000room}, and more recently acknowledged in the EMC community \cite{Corona2002}. Nevertheless, as our purpose is to address the physical situations encountered in reverberation chambers, we will start from a parallelepipedic cavity and introduce step by step simple modifications of the boundary. The dimensions of the bare cavity are length: 0.985\,m, width: 0.785\,m, height: 0.995\,m. We begin by adding a single hemisphere of radius 0.15\,m on the ceiling of the chamber (see Fig. \ref{schema}). The chaotic cavity thus obtained is a 3D realization of a dispersing billiard, the well-known Sinai billiard \cite{Alt1997,Dorr1998}.

\begin{figure}[h]
 \resizebox{\columnwidth}{!}{\includegraphics{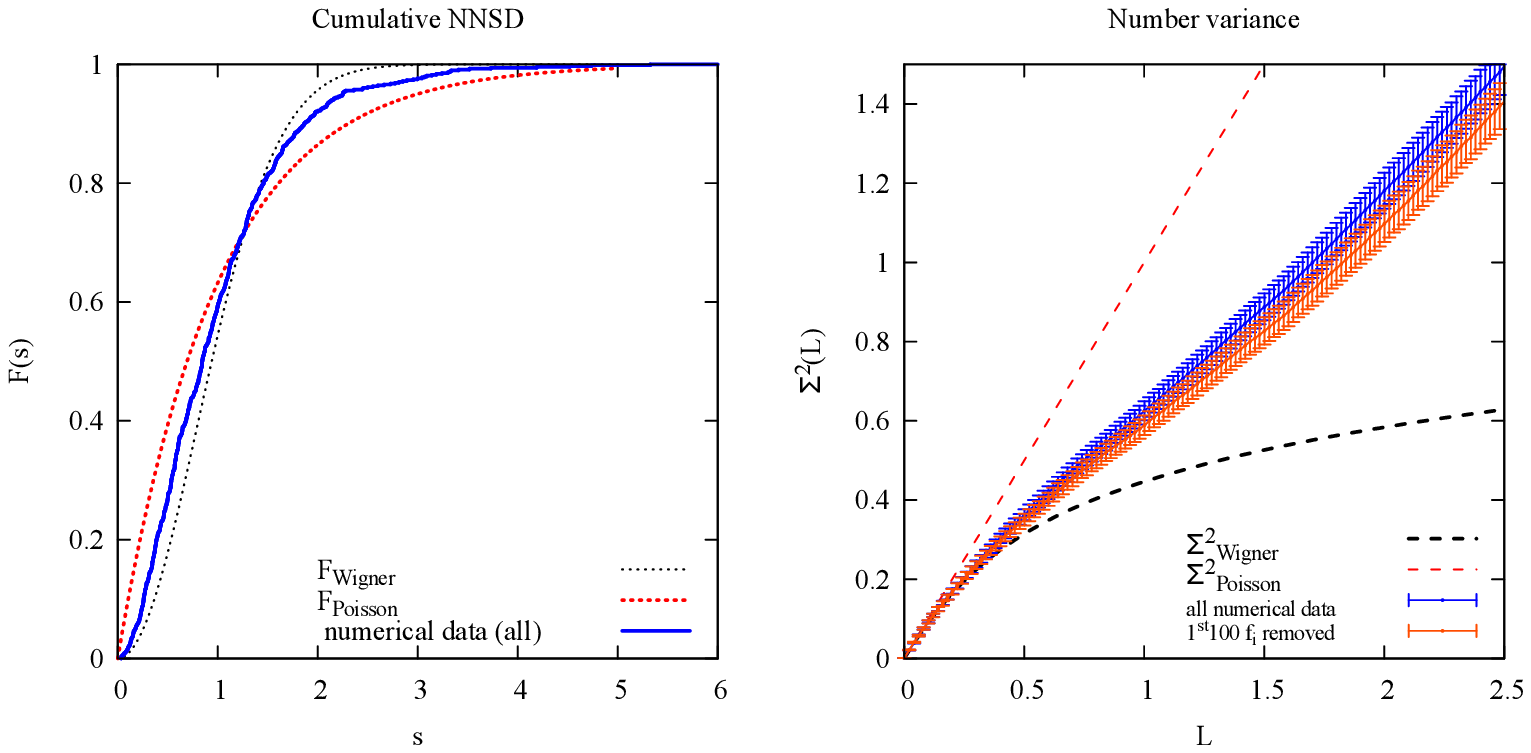}}
 \caption{Left: Cumulative distribution function $F(s)$ for the first $681$ modes of the cavity shown in Fig. \ref{schema}. The numerical data lies between the cumulative distributions associated to $ P_\textrm{Poisson}$ and $P_\textrm{Wigner}$.
 By fitting with $ P_\textrm{Weibull}$ one gets $\beta=1.45$. Right: Number variance for the same data. Even by omitting the first hundred modes (orange data) the agreement with $\Sigma^2_\textrm{Wigner}$ is barely improved.}
 \label{sinai}
\end{figure}

In this cavity, we used the commercial software Comsol\textregistered\  to compute the first 681 modes with perfectly conducting walls. In Fig. \ref{sinai}, the corresponding cumulative distribution function $F(s)$ (\ref{cdf}) and the number variance (\ref{sigma2}) are plotted. Both show significant deviations from GOE predictions indicating that the universal regime, where ergodic modes are dominating, is not reached within this range of frequency. Indeed, asymptotically, for wavelengths much smaller than the sphere radius, it is well established that a Sinai-like cavity exhibits GOE spectral fluctuations \cite{Alt1997}. By fitting with $P_\textrm{Weibull}$, we obtain $\beta=1.45$. Even by omitting the first 100 modes, the majority of which behave very closely to modes of the bare cavity, one still obtains $\beta=1.50$. It is worth noting that we obtain similar results in simulations of a mode-stirred chamber for a given position of the stirrer in the same frequency range \cite{suivant}. In this case, we observe that the asymptotic spectral behaviour is intermediate between Poisson and GOE. In 2D cavities, similar \emph{semi-Poisson} statistics  are obtained either with a barrier \cite{Wiersig2002} or a pointlike scatterer \cite{Laurent2006} (i.e. in pseudo-integrable cavities). However, for the data presented in Fig. \ref{sinai}, semi-Poisson could not reproduce the behaviour of $F(s)$ for small and large $s$ as satisfactorily as Weibull.

\begin{figure}[h!]
\center
\resizebox{16cm}{6cm}{\includegraphics{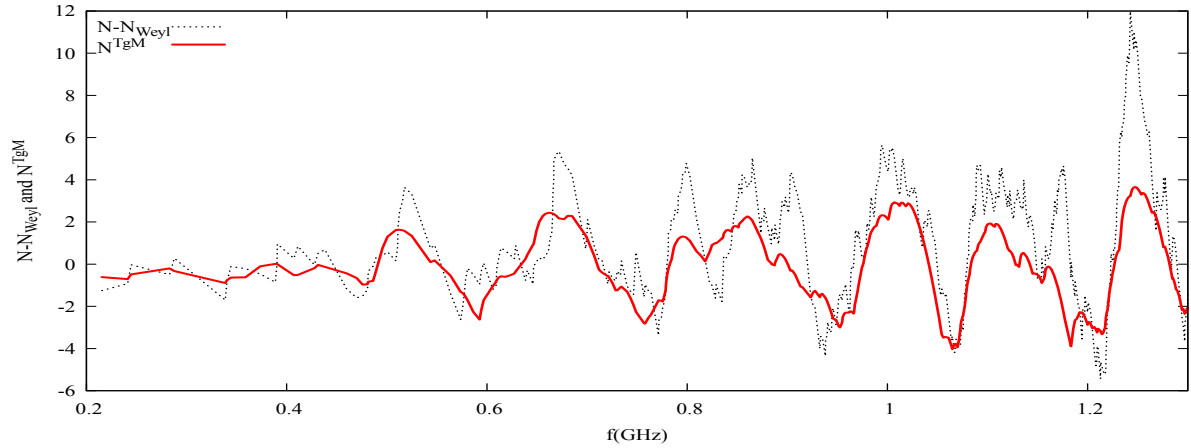}}
\caption{Dotted black curve: $N-\Nw$, red curve $N^{TgM}$ : Contribution of TgM to fluctuations of $N-\Nw$. Important large  scale fluctuations due to a statistically important fraction of non generic modes in the spectrum can be observed }
\label{Ntgm}
\end{figure}

The deviations shown in Fig. \ref{sinai} are due to the fact that, in spite of the presence of the half-sphere, the chamber still exhibits regular features (parallel walls). The most important family of \emph{regular} modes consists of tangential modes (TgM) which, in the bare cavity, have a single null wave vector component. The non zero components are quantised in the planes parallel to the walls of the cavity as shown in Fig. \ref{schema}. The counting function for TgM perpendicular to the $z$-axis is then obtained by following the procedure introduced in \cite{Dembowski2002}:
 \begin{align}
  N^z_t(k)&=\sum_{\mu,\nu}\int \frac{dz}{2\pi}\int dk_z \Theta(k^2-k^2_{x\mu}-k^2_{y\nu}-k^2_z)  \label{coutingtgm}\\ 
&\textrm{with } k_{x\mu}=\mu\pi/L_x,k_{y\nu}=\nu\pi/L_y\\
 N^z_t(k)&=\lambda_z \frac{L_z}{\pi} \sum_{k^2 \geq 
k^2_{x\mu}+k^2_{y\nu}}\sqrt{k^2-k^2_{x\mu}-k^2_{y\nu}}^{\textbf{-}} 
 \end{align}
where $L_x$ and $L_y$ are  the dimension of the bare cavity in $x$ and $y$ directions respectively and  $\lambda_z$ is an adjustable parameter accounting for the presence of the sphere which limits the extension along $z$. The corresponding expressions for $N^x_t$ and $N^y_t$ are simply deduced from circular permutations of the indices.  The total counting function of TgM, denoted $N_t=N^x_t+N^y_t+N^z_t$, has a smoothly varying part $ N_t^\textrm{av}$ which can be fitted by a polynomial function of order 3. The fluctuations of $N_t$ around $ N_t^\textrm{av}$, denoted by $N^{TgM}=N_t-N_t^\textrm{av}$,  and those of $N$ around Weyl's prediction are compared in Fig. \ref{Ntgm}. The comparison clearly shows that the tangential modes  are  responsible for the wide large scale fluctuations of the counting function. Thus, in this range of frequency, 
these regular modes (due to the remaining parallel walls) are not so scarce and lead to strong deviations from GOE behaviour.

\begin{figure}[h!]
\resizebox{\columnwidth}{!}{\includegraphics{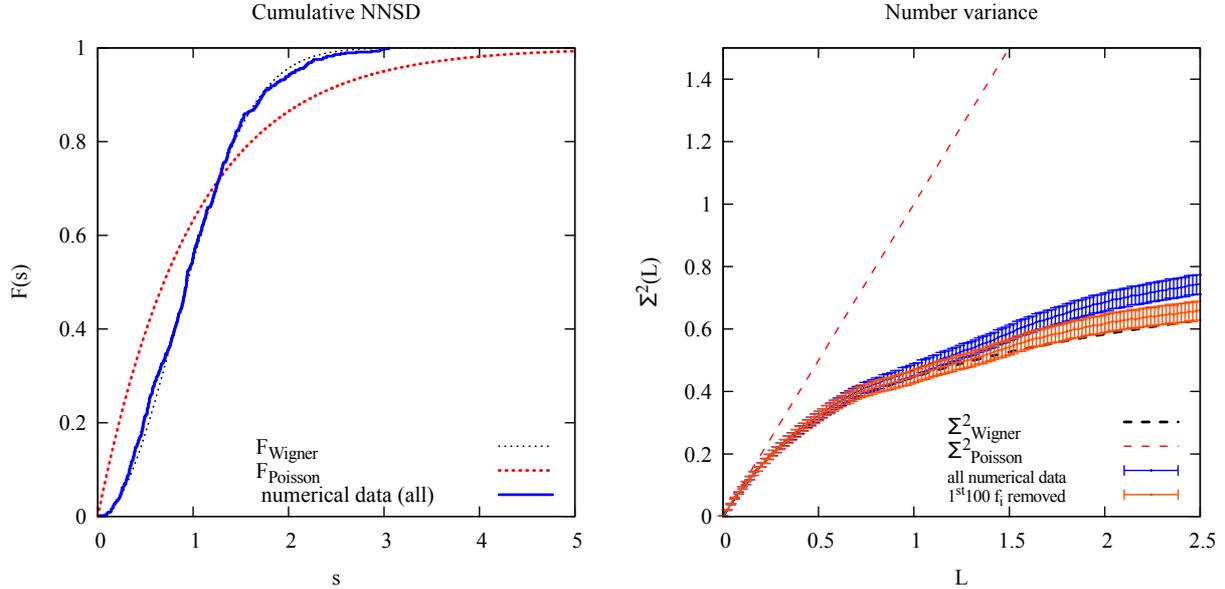}}
\caption{Cumulative NNSD and number variance for  the first 571 modes of the cavity with 1 half-sphere and 2 caps shown in Fig. \ref{statmode}. The results obtained  are very close to those expected for a chaotic cavity in the universal regime.}
\label{3caps}
\end{figure}

\begin{figure}[h!]
 \resizebox{\columnwidth}{!}{\includegraphics{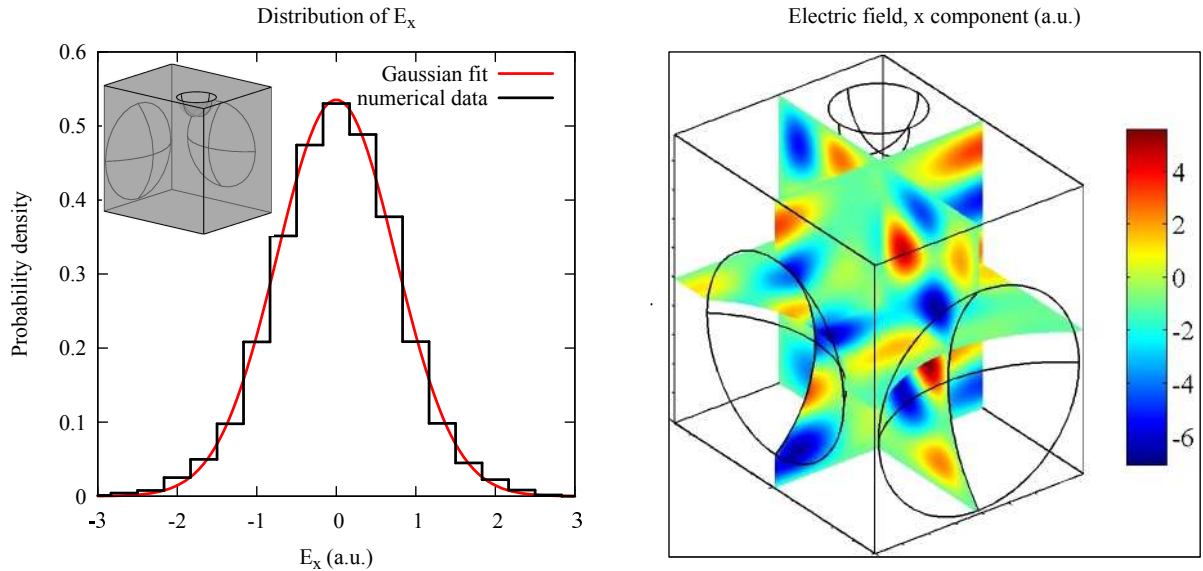}}
\caption{Left: Probability density of the $x-$component of the electric field for the $230^\textrm{th}$ mode at $1.03$ GHz of the cavity shown in insert  (Gaussian fit in red). Right: color map of the same component in three orthogonal plane.The results for the other components are similar.}
 \label{statmode}
\end{figure}

The previous results indicate that adding a half-sphere is not a sufficient modification of the parallelepipedic RC to reach the universal properties of a wave chaotic cavity, especially when dealing with the first hundreds modes. A major improvement of the geometry is then performed by introducing in the chamber two spherical caps with respective radii of 55\,cm and 45\,cm. As shown in Fig.\ref{statmode}, both caps are not centred and penetrate inside the cavity to a maximum length of $15$\,cm. Thus, the usable volume of the cavity is not significantly reduced. The first 571 eigenmodes of this chamber have been computed following the same numerical procedure as previously used. In Fig. \ref{3caps} the cumulative function $F(s)$ and the number variance $\Sigma^2$ of the new spectrum are shown. The results obtained  are very close to those expected for a chaotic cavity in the universal regime ; the Weibull parameter $\beta$ reaches now $\beta=1.86$, ie $\beta=2-2\sigma_\beta$, $F(s)$ is almost indistinguishable from $F_\textrm{Wigner}$. Compared to the RC with the single half-sphere (Fig. \ref{sinai}), the behaviour of $\Sigma^2$ (blue curve) is extremely improved. The agreement with $\Sigma^2_{GOE}$ is almost perfect (orange  curve) when omitting the first hundred modes, in which case $\beta=1.97$. These results show that the cavity obtained with the half-sphere and the two caps is an excellent realisation of a chaotic RC.

The Fig. \ref{statmode} shows  a typical mode of the chaotic RC and the corresponding amplitude distribution for the $x-$component of the electric field. As stressed in \cite{wright2010_4}, such statistics are spoiled when the field close to the walls is taken into account. Thus, we performed the analysis by excluding a zone $\lambda/4$ wide at the boundaries. Its spectral statistics being given by GOE, the eigenmodes of this cavity are expected to be \emph{ergodic}, with Gaussian spatial distributions. This behaviour is clearly observed in Fig. \ref{statmode}. Except for the lowest modes, the same ergodic properties are obtained for all the Cartesian components of almost  all the modes.

\section{Response in the presence of losses: towards a reduced LUF in chaotic RCs}

In all the cavities studied above, the knowledge of the modes for perfectly conducting walls is not enough to account for the actual response in such cavities due to the presence of unavoidable sources of losses. These losses are related to the finite conductivity of the metals used for the boundaries of the RC and also to all possible interactions with electric devices such as antennas.
The latter are indeed known to bring the dominant contribution to losses in an RC operating at low and moderate frequencies (typically the first few hundred modes).
A simple way to qualify a well operating RC by taking perturbative losses into account is to calculate the response of the cavity when it is excited by an elementary oscillating dipole source located at $\vec{r}_0 $ at frequency $f_0$  through the dyadic Green's function (DGF) $\overline{\overline {G}}(\vec{r},\vec{r}_0,f_0)$.
Far from the source region, the DGF can be expanded over the resonances following the formalism given in \citep{dolmans1995electromagnetic}:
 \begin{equation}
  \overline{\overline {G}}(\vec{r},\vec{r}_0,f_0)=\sum_{n=1}^{\infty} \frac{\tilde{k}_n^2 \vec{E}_n(\vec{r})\vec{E}_n(\vec{r_0})} {k_0^2(k_n^2-k_0^2)}\textrm{,where }k_0=2\pi f_0 / c
  \label{a}
 \end{equation}
Note that certain terms are omitted in the above expression, which may become important when the source and the observation point are very close together \citep{dolmans1995electromagnetic}. In  expression (\ref{a}), losses lead to complex eigenmodes $\vec{E}_n$ (\citep{Barthelemy2003,Barthelemy2005,Poli2010}) associated to complex eigenvalues $k_n$ which can be written: 
\begin{equation}
 k_n=\tilde{k}_n(1-\frac{1+i}{2 Q_n})
 \label{b}
 \end{equation}
 where  $\tilde{k}_n=2\pi f_n/c$ is the eigenvalue of the lossless  cavity and $Q_n$ is the quality factor of the n$^\textrm{th}$ resonance related to the n$^\textrm{th}$ resonance bandwidth $\Delta_{f_n}=\tilde{k}_n/(2 \pi Q_n)$.
  
An important parameter for this approach is the modal overlap at $f_0$: 
  \begin{equation}
M(f_0)=\frac{\langle\Delta_{k_n}\rangle}{\Delta(k_0)}=\frac{8\pi V}{c^3}\frac{f_0^3}{Q}
 \label{c}
 \end{equation}
 where $\langle\Delta_{k_n}\rangle$ is the mean resonance bandwidth, $\Delta(k_0)$ is the mean spectral spacing at $k_0$, and $Q$ is the average quality factor.
% \end{itemize}  

From previous studies \cite{Savin2006,Poli2009,Poli2010}, it has been shown that complex eigenmodes are due to the presence of traveling waves superimposed to standing waves. In a perturbative approach,  the statistics of resonance states in a two-dimensional chaotic microwave cavity were investigated by solving the Maxwell equations with lossy boundaries subject to Ohmic dissipation. A successful comparison of the statistics of its complex-valued resonance states and associated bandwidths with analytical predictions based on a non-Hermitian effective Hamiltonian model was achieved. As shown in \citep{Poli2010}, a weakly coupled regime corresponds to fluctuations of the widths which are small compared to the mean spacing. In the following, we will therefore assume that, in a given restricted frequency band, all the bandwidths are equal to their mean value and that the imaginary part of the field components of the eigenmodes can be neglected, so that the $\vec{E}_n$ in (\ref{a}) are those of the corresponding lossless cavity.

\begin{figure}[h!]\resizebox{\columnwidth}{5.5cm}{\includegraphics{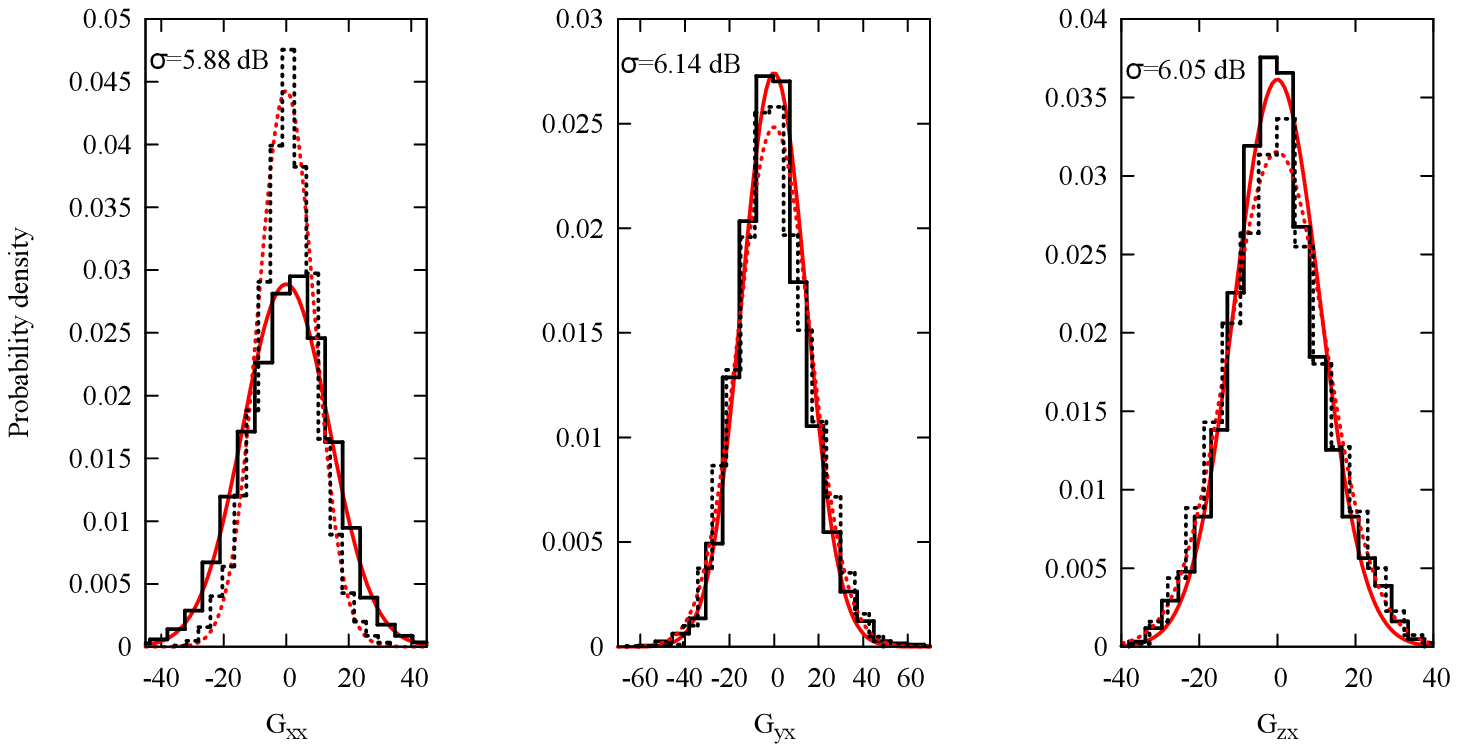}}
\resizebox{\columnwidth}{5.5cm}{\includegraphics{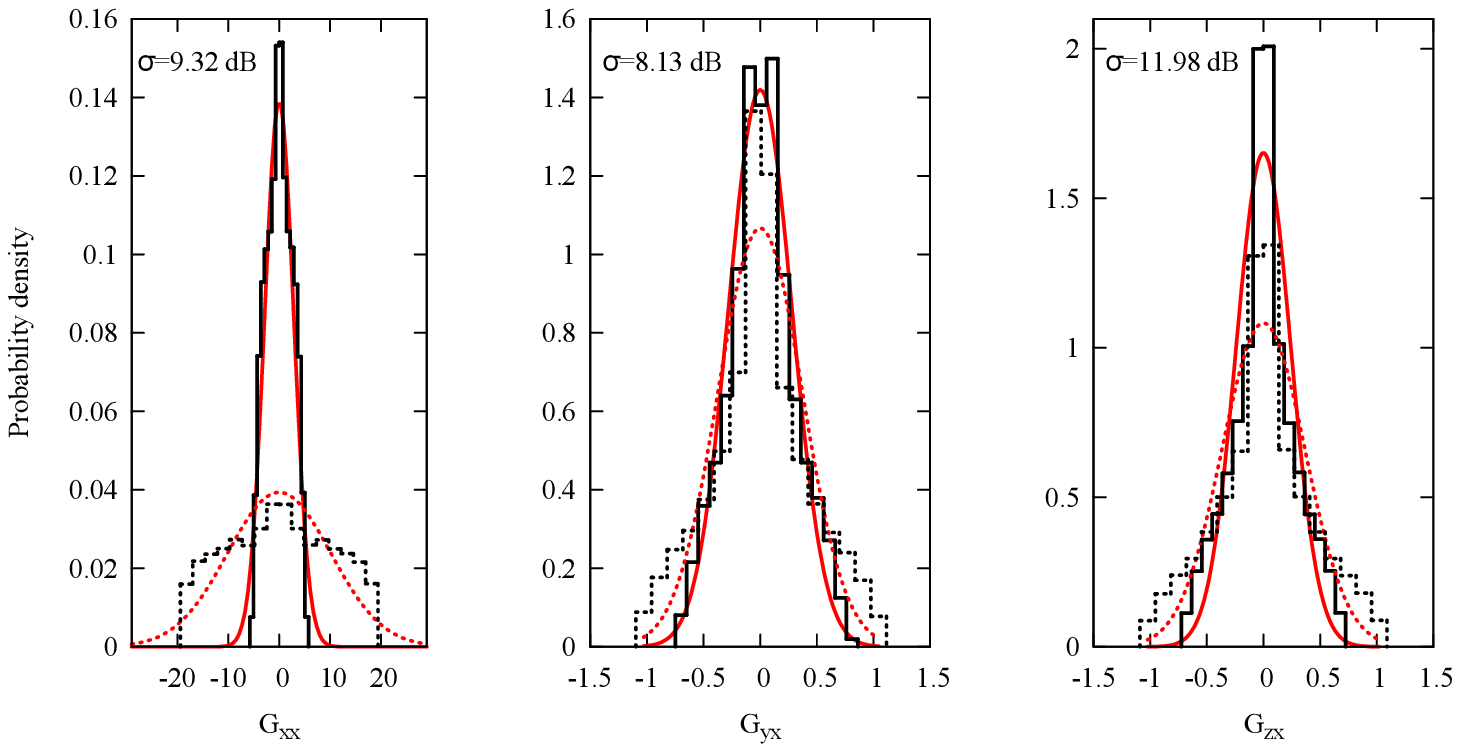}}
\caption{Spatial distribution of the responses for an excitation polarized along the $x$-axis at a frequency $f_0=953.6$ MHz:  in the parallelepipedic cavity (bottom); in the chaotic cavity shown in Fig. \ref{statmode} (top).
 All numerical distributions are fitted by a Gaussian distribution (in red). Continuous lines are used for the real part and dotted lines for the imaginary part. The standard deviations $\sigma$ (in dB) of the power density (see text) are given as inserts. }
\label{green_3c_rec}
\end{figure}

In the following, we will compare the responses of the chaotic cavity shown in Fig. \ref{statmode} and of a bare cavity, both with the same volume $V=0.6996$ m$^3$. In order to account for realistic values of $Q$, we used numerical and experimental data obtained in different RCs with comparable sizes \citep{CAOREV}. Essentially, the global quality factor is given by $ 1/Q=1/Q_\textrm{walls}+1/Q_\textrm{antennas}$. At frequencies near the LUF $\simeq 3$ to $6 f_c$ (corresponding to the interval $ \left[ 620 \,\textrm{MHz}, 1240 \,\textrm{MHz}\right]$ in the cavities studied below), the expected values for $Q_\textrm{walls}$ are of the order of $10^5$, whereas  $Q_\textrm{antennas}$  typically ranges between $10^3$ -- $10^4$. Therefore $Q$ is dominated by the losses due to the antennas. This motivated us to fix the value of $Q$ to $10^3$ in the computations of the DGFs. In Fig. \ref{green_3c_rec} we consider the responses for an excitation polarized along the $x$-axis at a frequency $f_0=953.6$ MHz ($\simeq 4.6 f_c$ in the chaotic cavity shown in Fig. \ref{statmode}). This choice corresponds to a moderately small modal overlap, $M(f_0)=0.57$, which is clearly not in the overmoded regime. In this Figure, numerical spatial distributions of the complex  field Cartesian components are shown. The Gaussian character of these distributions is clearly established in the case of the chaotic cavity. It is particularly interesting  to remark that the real and imaginary parts of each field component have different standard deviations,  this fact being obviously in contradiction with the assumption of a continuous plane wave-spectrum \cite{Hill1998,hill2009electromagnetic}. 
For each shown component, the standard deviations of the power density ($ 10 \log (\modulc{G_{ab}}{}/ \modulc{G_\textrm{ref}}{})$ in dB) are given as inserts in Fig. \ref{green_3c_rec}. Note that the values are close for the chaotic cavity and are much more scattered in the bare cavity. These values are to be compared with those measured by Mitra and Trost \cite{Mitra} in a RC with a comparable volume. The authors suggest that the standard deviation values should be comprised between $5$ and $6$ dB for a well-stirred operation, which are typically obtained for values of the modal overlap between $1.0$ to $2.5$, corresponding to frequencies well above the frequency $f_0$ considered in Fig. \ref{green_3c_rec}.
Thus, it is remarkable that the main statistical requirements of a well-operating RC, namely the homogeneity and the isotropy of the field distribution, can be obtained  without the help of any stirring process in a chaotic cavity.

\section{Conclusion}
In this paper, by a step by step modification of a regular RC, we demonstrated that a 3D chaotic cavity can display universal spatial and spectral statistics in the low frequency range provided that the geometry is such that all regular modes (mostly \emph{tangential modes}) are suppressed. Moreover we have shown how the response of a 3D fully chaotic  RC, accounting for the presence of losses, can fulfil all the statistical requirements of a well-operating  RC, concerning the spatial distribution of the field, without the use of any spatial stirring. We also suggest that this type of improvement for a RC can significantly reduce the lowest usable frequency of operation.

\section*{Acknowledgements}
The authors thank  Odile Picon for fruitful discussions and the french National Research Agency (ANR) for financially supporting the project CAOREV of which this work is a part.

\clearpage

\bibliographystyle{elsarticle-num}
\bibliography{inno_v2}

\begin{thebibliography}{10}
\expandafter\ifx\csname url\endcsname\relax
  \def\url#1{\texttt{#1}}\fi
\expandafter\ifx\csname urlprefix\endcsname\relax\def\urlprefix{URL }\fi
\expandafter\ifx\csname href\endcsname\relax
  \def\href#1#2{#2} \def\path#1{#1}\fi

\bibitem{Norme2003}
{CISPR/A and IEC SC 77B, IEC 61000-4-21 – Electromagnetic Compatibility (EMC)
  - Part 4-21: Testing and Measurement Techniques - Reverberation Cham- ber
  Test Methods, International Electrotechnical Commission (IEC) International
  standard}.

\bibitem{Bruns2005}
C.~Bruns,
  \href{http://e-collection.library.ethz.ch/eserv/eth:28055/eth-28055-02.pdf}{{Three-dimensional
  simulation and experimental verification of a reverberation chamber}}, Ph.D.
  thesis (2005).
\newline\urlprefix\url{http://e-collection.library.ethz.ch/eserv/eth:28055/eth-28055-02.pdf}

\bibitem{Hill1998a}
D.~Hill, {Electromagnetic theory of reverberation chambers}, NIST Technical
  note 1506.

\bibitem{hill2009electromagnetic}
D.~Hill, {Electromagnetic fields in cavities: deterministic and statistical
  theories}, IEEE Press Series on Electromagnetic Wave Theory, IEEE ; Wiley,
  2009.

\bibitem{Cozza2011}
A.~Cozza, {The Role of Losses in the Definition of the Overmoded Condition for
  Reverberation Chambers and Their Statistics}, IEEE Transactions on
  Electromagnetic Compatibility 53~(2) (2011) 296--307.

\bibitem{Hill1998}
D.~Hill, {Plane wave integral representation for fields in reverberation
  chambers}, IEEE Transactions on Electromagnetic Compatibility 40~(3) (1998)
  209--217.

\bibitem{Arnaut2001}
L.~Arnaut, {Operation of electromagnetic reverberation chambers with wave
  diffractors at relatively low frequencies}, IEEE Transactions on
  Electromagnetic Compatibility 43~(4) (2001) 637--653.

\bibitem{Lunden2007}
O.~Lund{\'e}n, M.~B{\"a}ckstr{\"o}m, How to avoid unstirred high frequency
  components in mode stirred reverberation chambers, in: Electromagnetic
  Compatibility, 2007. EMC 2007. IEEE International Symposium on, 2007, pp.
  1--4.
\newblock \href {http://dx.doi.org/10.1109/ISEMC.2007.244}
  {\path{doi:10.1109/ISEMC.2007.244}}.

\bibitem{Hong2010}
J.-I. Hong, C.-S. Huh, Optimization of stirrer with various parameters in
  reverberation chamber, Progress In Electromagnetics Research 104 (2010)
  15--30.

\bibitem{Berry1977}
M.~V. Berry, {Regular and irregular semiclassical wavefunctions}, Journal of
  Physics A: Mathematical and General 10~(12) (1977) 2083.

\bibitem{wright2010_2}
O.~Legrand, F.~Mortessagne, {Wave Chaos for the Helmholtz Equation}, New
  Directions in Linear Acoustics and Vibration: Quantum Chaos, Random Matrix
  Theory, and Complexity, Cambridge University Press, 2010.

\bibitem{wright2010_4}
M.~Dennis, {Gaussian RandomWavefields and the Ergodic Mode Hypothesis}, New
  Directions in Linear Acoustics and Vibration: Quantum Chaos, Random Matrix
  Theory, and Complexity, Cambridge University Press, 2010.

\bibitem{BGS}
O.~Bohigas, M.~Giannoni, C.~Schmit, {Characterization of chaotic quantum
  spectra and universality of level fluctuation laws}, Physical Review Letters
  52~(1) (1984) 1--4.

\bibitem{Bohigas1984}
O.~Bohigas, M.-J. Giannoni, {Chaotic motion and random matrix theories}, in:
  J.~S. Dehesa, J.~M.~G. Gomez, A.~Polls (Eds.), Mathematical and Computational
  Methods in Nuclear Physics, Vol. 209 of Lecture Notes in Physics, Springer
  Berlin Heidelberg, 1984, pp. 1--99.

\bibitem{stockmann1999quantum}
H.~J. St\"{o}ckmann, {Quantum Chaos: An Introduction}, Cambridge University
  Press, 1999.

\bibitem{McDonald1988}
S.~McDonald, A.~Kaufman, {Wave chaos in the stadium: statistical properties of
  short-wave solutions of the Helmholtz equation}, Physical Review A 37~(8)
  (1988) 3067--3086.

\bibitem{Stein1992}
J.~Stein, H.-J. St\"{o}ckmann, {Experimental determination of billiard wave
  functions}, Physical Review Letters 68~(19) (1992) 2867--2870.

\bibitem{Alt1994}
H.~Alt, H.-D. Gr\"{a}f, H.~Harney, R.~Hofferbert, {Superconducting billiard
  cavities with chaotic dynamics: An experimental test of statistical
  measures}, Physical Review E 50~(1) (1994) R1--R4.

\bibitem{Pre_Doya}
V.~Doya, O.~Legrand, F.~Mortessagne, C.~Miniatura, {Speckle statistics in a
  chaotic multimode fiber}, Physical Review E 65~(5) (2002) 056223.

\bibitem{Deus1995}
S.~Deus, P.~M. Koch, L.~Sirko, {Statistical properties of the eigenfrequency
  distribution of three-dimensional microwave cavities}, Physical Review E
  52~(1) (1995) 1146--1155.

\bibitem{Alt1997}
H.~Alt, C.~Dembowski, H.~Gr\"{a}f, R.~Hofferbert, {Wave dynamical chaos in a
  superconducting three-dimensional sinai billiard}, Physical Review Letters 79
  (1997) 1026--1029.

\bibitem{Dembowski2002}
C.~Dembowski, B.~Dietz, H.~D. Graef, A.~Heine, T.~Papenbrock, A.~Richter,
  C.~Richter, {Experimental Test of a Trace Formula for a Chaotic Three
  Dimensional Microwave Cavity}, Physical Review Letters 89~(6) (2002) 4.
\newblock \href {http://arxiv.org/abs/0206028} {\path{arXiv:0206028}}.

\bibitem{Dietz2008}
B.~Dietz, B.~M\"{o}{\ss}ner, T.~Papenbrock, U.~Reif, A.~Richter, {Bouncing ball
  orbits and symmetry breaking effects in a three-dimensional chaotic
  billiard}, Physical Review E 77~(4 Pt 2) (2008) 11.
\newblock \href {http://arxiv.org/abs/0804.1551} {\path{arXiv:0804.1551}}.

\bibitem{Dorr1998}
U.~D\"{o}rr, H.-J. St\"{o}ckmann, M.~Barth, U.~Kuhl, {Scarred and chaotic field
  distributions in a three-dimensional Sinai-microwave resonator}, Physical
  Review Letters 80~(5) (1998) 1030--1033.

\bibitem{Orjubin2009}
G.~Orjubin, E.~Richalot, O.~Picon, O.~Legrand, {Wave chaos techniques to
  analyze a modeled reverberation chamber}, Comptes Rendus Physique 10~(1)
  (2009) 42--53.

\bibitem{Gradoni2011}
G.~Gradoni, J.-H. Yeh, T.~M. Antonsen, S.~Anlage, E.~Ott, {Wave chaotic
  analysis of weakly coupled reverberation chambers}, in: 2011 IEEE
  International Symposium on Electromagnetic Compatibility, Institute for
  Research in Electronics and Applied Physics, University of Maryland, College
  Park 20742, USA, IEEE, 2011, pp. 202--207.

\bibitem{Hannay1999}
J.~Hannay, R.~McCraw, {Barrier billiards-a simple pseudointegrable system},
  Journal of Physics A: Mathematical and General 23~(6) (1999) 887--900.

\bibitem{Bogomolny2004}
E.~Bogomolny, C.~Schmit, {Structure of Wave Functions of Pseudointegrable
  Billiards}, Physical Review Letters 92~(24) (2004) 244102.

\bibitem{Bogomolny2006}
E.~Bogomolny, B.~Dietz, T.~Friedrich, M.~Miski-Oglu, A.~Richter,
  F.~Sch\"{a}fer, C.~Schmit, {First Experimental Observation of Superscars in a
  Pseudointegrable Barrier Billiard}, Physical Review Letters 97~(25) (2006)
  254102.

\bibitem{Luk}
W.~Lukosz, {Electromagnetic zero-point energy shift induced by conducting
  surfaces}, Zeitschrift f{\"u}r Physik 262~(4) (1973) 327--348.

\bibitem{Balian1977}
R.~Balian, B.~Duplantier, {Electromagnetic waves near perfect conductors. I.
  Multiple scattering expansions. Distribution of modes}, Annals of physics
  104~(2) (1977) 300--335.

\bibitem{Orjubin2007}
G.~Orjubin, E.~Richalot, O.~Picon, O.~Legrand, {Chaoticity of a Reverberation
  Chamber Assessed From the Analysis of Modal Distributions Obtained by FEM},
  IEEE Transactions on Electromagnetic Compatibility 49~(4) (2007) 762--771.

\bibitem{Tanner1997}
G.~Tanner, {How chaotic is the stadium billiard? A semiclassical analysis},
  Journal of Physics A: Mathematical and General 30 (1997) 2863--2888.

\bibitem{Backer1997}
A.~B{\"a}cker, R.~Schubert, P.~Stifter, {On the number of bouncing ball modes
  in billiards}, Journal of Physics A: Mathematical and General 30 (1997)
  6783--6795.

\bibitem{Bunimovich1998}
L.~A. Bunimovich, J.~Rehacek, \href{http://eudml.org/doc/76791}{On the
  ergodicity of many-dimensional focusing billiards}, Annales de l'institut
  Henri Poincar{\'e} (A) Physique th{\'e}orique 68~(4) (1998) 421--448.
\newline\urlprefix\url{http://eudml.org/doc/76791}

\bibitem{Berry1981}
M.~V. Berry, Regularity and chaos in classical mechanics, illustrated by three
  deformations of a circular billiard, European Journal of Physics 2 (1981)
  91--102.

\bibitem{kuttruff2000room}
H.~Kuttruff, Room acoustics, Taylor \& Francis, 2000.

\bibitem{Corona2002}
P.~Corona, J.~Ladbury, G.~Latmiral, Reverberation-chamber research-then and
  now: a review of early work and comparison with current understanding,
  Electromagnetic Compatibility, IEEE Transactions on 44~(1) (2002) 87--94.

\bibitem{suivant}
J.-B. Gros, O.~Legrand, F.~Mortessagne, O.~Picon, E.~Richalot, K.~Selemani, in
  preparation.

\bibitem{Wiersig2002}
J.~Wiersig, {Spectral properties of quantized barrier billiards}, Physical
  Review E 65~(4) (2002) 046217.

\bibitem{Laurent2006}
D.~Laurent, O.~Legrand, F.~Mortessagne, {Diffractive orbits in the length
  spectrum of a two-dimensional microwave cavity with a small scatterer},
  Physical Review E 74~(4) (2006) 046219.

\bibitem{dolmans1995electromagnetic}
G.~Dolmans,
  \href{http://alexandria.tue.nl/extra1/erap/publichtml/9510769.pdf}{{Electromagnetic
  Fields Inside a Large Room with Perfectly Conducting Walls}}, EUT Report E,
  Eindhoven University of Technology, Faculty of Electrical Engineering, 1995.
\newline\urlprefix\url{http://alexandria.tue.nl/extra1/erap/publichtml/9510769.pdf}

\bibitem{Barthelemy2003}
J.~Barth\'{e}lemy, \href{http://tel.archives-ouvertes.fr/tel-00004114}{{Chaos
  ondulatoire en pr\'{e}sence de pertes: mod\'{e}lisation et exp\'{e}rience de
  billards micro-ondes}}, Ph.D. thesis, Nice Sophia Antipolis (2003).
\newline\urlprefix\url{http://tel.archives-ouvertes.fr/tel-00004114}

\bibitem{Barthelemy2005}
J.~Barth\'{e}lemy, O.~Legrand, F.~Mortessagne, {Complete S matrix in a
  microwave cavity at room temperature}, Physical Review E 71~(1) (2005)
  016205.

\bibitem{Poli2010}
C.~Poli, O.~Legrand, F.~Mortessagne, {Statistics of resonance states in a
  weakly open chaotic cavity with continuously distributed losses}, Physical
  Review E 82~(5) (2010) 055201.

\bibitem{Savin2006}
D.~V. Savin, O.~Legrand, F.~Mortessagne, {Inhomogeneous losses and complexness
  of wave functions in chaotic cavities}, Europhysics Letters (EPL) 76~(5)
  (2006) 774--779.

\bibitem{Poli2009}
C.~Poli, D.~V. Savin, O.~Legrand, F.~Mortessagne, {Statistics of resonance
  states in open chaotic systems: A perturbative approach}, Physical Review E
  80~(4) (2009) 046203.

\bibitem{CAOREV}
J.-B. Gros, O.~Legrand, F.~Mortessagne, O.~Picon, E.~Richalot, K.~Selemani,
  {ANR CAOREV} report.

\bibitem{Mitra}
A.~K. Mitra, T.~F. Trost, Statistical simulations and measurements inside a
  microwave reverberation chamber, in: Electromagnetic Compatibility, 1997.
  IEEE 1997 International Symposium on, 1997, pp. 48--53.

\end{thebibliography}

\end{document}